\documentstyle[preprint,aps]{revtex}

\begin{document}
\tightenlines
\draft

\preprint{\vbox{\hbox{SLAC-PUB-6345}
                \hbox{WIS-93/86/Sep-PH}
                \hbox{UMHEP-382}
                \hbox{September 1993}
                \hbox{T/E} }}
\vskip 0.5truecm

\title{The Decay $B\to\pi\,\ell\,\nu$ in Heavy Quark Effective Theory}

\author{ Gustavo Burdman$^a$, Zoltan Ligeti$^b$, Matthias Neubert$^c$,
and Yosef Nir$^b$ }

\address{ \vbox{\vskip 0.5truecm}
  $^a$Department of Physics and Astronomy, \\
      University of Massachusetts, Amherst, MA 01003 \\
  \vbox{\vskip 0.2truecm}
  $^b$Weizmann Institute of Science,  \\
      Physics Department, Rehovot 76100, Israel \\
  \vbox{\vskip 0.2truecm}
  $^c$Stanford Linear Accelerator Center,  \\
      Stanford University, Stanford, CA 94309 }

\maketitle

\begin{abstract}
We present a systematic analysis of the $B^{(*)}\to\pi\,\ell\,\nu$ weak
decay form factors to order $1/m_b$ in the heavy quark effective
theory, including a discussion of renormalization group effects. These
processes are described by a set of ten universal functions (two at
leading order, and eight at order $1/m_b$), which are defined in terms
of matrix elements of operators in the effective theory. In the soft
pion limit, the effective theory yields normalization conditions for
these functions, which generalize the well-known current algebra
relations derived from the combination of heavy quark and chiral
symmetries to next-to-leading order in $1/m_b$. In particular, the
effects of the nearby $B^*$-pole are correctly contained in the form
factors of the effective theory. We discuss the prospects for a model
independent determination of $|V_{ub}|$ and the $B B^*\pi$ coupling
constant from these processes.

\end{abstract}

%\pacs{}

\vskip 0.5truecm
\centerline{(submitted to Physical Review D)}
\newpage

\narrowtext
\section{Introduction}

Of the three independent mixing angles in the
Cabibbo--Kobayashi--Maskawa matrix, $|V_{ub}|$ is the most poorly
determined. Chiral symmetry provides an absolute normalization of the
hadronic form factor in the decay $K\to\pi\,\ell\,\nu$, allowing a
precise and model independent determination of $|V_{us}|$ \cite{Leut}.
Heavy quark symmetry provides absolute normalization and various
relations among the form factors in the decays $B\to D^{(*)}\ell\,\nu$,
allowing a precise and model independent determination of $|V_{cb}|$
\cite{Shur,Nuss,Volo,Isgu,Vcb}. Neither of these symmetries is as
powerful in heavy-to-light transitions such as $b\to u\,\ell\,\bar\nu$.
Consequently, the present determination of $|V_{ub}|$ from the endpoint
of the lepton spectrum in semileptonic $B$ decays suffers from large
theoretical uncertainties and strong model dependence \cite{CLEO,ARGUS}.

It was suggested that the exclusive semileptonic decay mode
$B\to\pi\,\ell\,\nu$ could be used for a more reliable determination of
$|V_{ub}|$ \cite{Vub}. The basis for this hope is the fact that, to
leading order in the heavy quark expansion and over a limited kinematic
range, the corresponding form factors are related to those of
$D\to\pi\,\ell\,\nu$ by heavy quark flavor symmetry. The applicability
of this idea depends, besides experimental considerations, on the
importance of symmetry-breaking corrections of order $1/m_Q$. For the
related case of leptonic decays of heavy mesons, there are indications
from lattice gauge theory \cite{latt1,latt2,latt3,latt4} and QCD sum
rule calculations \cite{MN,BBBD,Elet,DoPa} that these power corrections
can be significant.

Our purpose in this study is to work out the structure of $1/m_b$
corrections for the $B^{(*)}\to\pi\,\ell\,\nu$ decay form factors using
the heavy quark effective theory. The main points of our analysis are
as follows:
\vspace{-8pt}
\begin{itemize}\itemsep=-4pt
\item[(i)]
Eight universal functions are needed to describe the $1/m_b$
corrections to these processes. They are defined in terms of matrix
elements of dimension four operators in the effective theory.
\item[(ii)]
The renormalization group improvement of these low-energy parameters
is discussed in detail.
\item[(iii)]
The behavior of the universal functions in the soft pion limit is
derived using standard current algebra techniques.
\item[(iv)]
It is shown explicitly that the $B^*$-pole contribution is correctly
contained in the heavy quark effective theory.
\end{itemize}
\vspace{-8pt}

At leading order in the heavy quark expansion, the two form factors
which parameterize $B\to\pi\,\ell\,\nu$ decays have been investigated
by several authors \cite{Vub,IWis,Wise,Burd,LW}. It is well-known that,
in this limit, the soft-pion behavior is fully determined by the decay
constant of the $B$ meson and the $B B^*\pi$ coupling constant. Here we
generalize these results to next-to-leading order in $1/m_b$. In
particular, we show that when one uses the physical meson decay
constants and the physical $B B^*\pi$ coupling constants (as opposed to
their asymptotic values in the $m_b\to\infty$ limit), there are neither
$1/m_b$ nor short-distance QCD corrections to the soft pion relations.
We also derive the general structure of the decay form factors at
larger pion momenta, where a chiral expansion is no longer valid.

The paper is organized as follows: The formalism of heavy quark
effective theory relevant to our work is reviewed in Sec.~\ref{sec:2}.
In Sec.~\ref{sec:3}, we then construct the heavy quark expansion for
the $B^{(*)}\to\pi\,\ell\,\nu$ decay form factors to next-to-leading
order in $1/m_b$, including a detailed analysis of renormalization
group effects. In Sec.~\ref{sec:4}, we derive the normalization
conditions for the universal functions of the effective theory, which
arise in the soft pion limit. We compare our results to the predictions
of the so-called heavy meson chiral perturbation theory
\cite{Wise,Burd}. Sec.~\ref{sec:5} contains a summary and some
concluding remarks concerning the prospects and possibilities to obtain
a model independent measurement of $|V_{ub}|$ and the $B B^*\pi$
coupling constant. Technical details related to the renormalization
group improvement and the soft pion limit are described in two
appendices.

\section{The $1/\lowercase{m}_Q$ Expansion}
\label{sec:2}

Our goal in this paper is to analyze the dependence of the hadronic
form factors describing $B\to\pi\,\ell\,\nu$ decays on the mass of the
$b$-quark, in the limit where $m_b\gg\Lambda_{\rm QCD}$. A convenient
tool to make this dependence explicit is provided by the heavy quark
effective theory (HQET)
\cite{Eich,Grin,Geor,Mann,Falk,Luke,FGL,AMM,review}. It is based on the
construction of an effective low-energy Lagrangian of QCD, which is
appropriate to describe the soft interactions of a heavy quark with
light degrees of freedom. In the effective theory, a heavy quark bound
inside a hadron moving at velocity $v$ is described by a
velocity-dependent field $h_v$, which is related to the conventional
quark field in QCD by \cite{Geor}
\begin{equation}\label{field}
   h_v(x) = \exp(i m_Q\,v\!\cdot\! x)\,{1+\rlap/v\over 2}\,Q(x) \,.
\end{equation}
By means of the phase redefinition one removes the large part of the
heavy quark momentum from the new field. When the total momentum is
written as $P=m_Q\,v+k$, the field $h_v$ carries the residual momentum
$k$, which results from soft interactions of the heavy quark with light
degrees of freedom and is typically of order $\Lambda_{\rm QCD}$. The
operator $\case{1}/{2}(1+\rlap/v)$ projects out the heavy quark (rather
than antiquark) components of the spinor. The antiquark components are
integrated out to obtain the effective Lagrangian
\cite{Eich,Geor,Mann,FGL}
\begin{equation}\label{lagr}
   {\cal{L}}_{\rm eff} = \bar h_v\,i v\!\cdot\!D\,h_v
   + {1\over 2 m_Q}\,\Big[ O_{\rm kin}
   + C_{\rm mag}(\mu)\,O_{\rm mag} \Big] + {\cal{O}}(1/m_Q^2) \,,
\end{equation}
where $D^\mu = \partial^\mu - i g_s T_a A_a^\mu$ is the gauge-covariant
derivative. The operators appearing at order $1/m_Q$ are
\begin{equation}\label{Omag}
   O_{\rm kin} = \bar h_v\,(i D)^2 h_v \,, \qquad
   O_{\rm mag} = {g_s\over 2}\,\bar h_v\,\sigma_{\mu\nu}
                 G^{\mu\nu} h_v \,.
\end{equation}
Here $G^{\mu\nu}$ is the gluon field strength tensor defined by
$[iD^\mu,iD^\nu]=i g_s G^{\mu\nu}$. In the hadron's rest frame, it is
readily seen that $O_{\rm kin}$ describes the kinetic energy resulting
from the residual motion of the heavy quark, whereas $O_{\rm mag}$
describes the chromo-magnetic coupling of the heavy quark spin to the
gluon field. One can show that, to all orders in perturbation theory,
the kinetic operator $O_{\rm kin}$ is not renormalized \cite{LuMa}. The
renormalization factor $C_{\rm mag}(\mu)$ of the chromo-magnetic
operator has been calculated in leading logarithmic approximation and
is given by \cite{FGL}
\begin{equation}\label{Cmag}
   C_{\rm mag} = x^{-3/\beta} \,, \qquad
   \beta = 11 - {2\over 3}\,n_f \,,
\end{equation}
where $x=\alpha(\mu)/\alpha(m_Q)$, $\mu$ denotes the renormalization
scale, and $n_f$ is the number of light quarks with mass below $m_Q$.

Any operator of the full theory that contains one or more heavy quark
fields can be matched onto a short-distance expansion in terms of
operators of the effective theory. In particular, the expansion of the
heavy-light vector current reads
\begin{equation}\label{current}
   \bar q\,\gamma^\mu\,Q \cong \sum_i C_i(\mu)\,J_i
   + {1\over 2 m_Q} \sum_j B_j(\mu)\,O_j + {\cal{O}}(1/m_Q^2) \,,
\end{equation}
where the symbol $\cong$ is used to indicate that this is an equation
that holds on the level of matrix elements. The operators $\{J_i\}$
form a complete set of local dimension-three current operators with the
same quantum numbers as the vector current in the full theory. In HQET
there are two such operators, namely
\begin{equation}\label{LOcurrent}
   J_1 = \bar q\,\gamma^\mu h_v \,, \qquad
   J_2 = \bar q\,v^\mu h_v \,.
\end{equation}
Similarly, $\{O_j\}$ denotes a complete set of local dimension-four
operators. It is convenient to use the background field method, which
ensures that there is no mixing between gauge-invariant and
gauge-dependent operators. Moreover, operators that vanish by the
equations of motion are irrelevant. It is thus sufficient to consider
gauge-invariant operators that do not vanish by the equations of
motion. A convenient basis of such operators is \cite{AMM}:
\begin{equation}\label{SLOcurrent}
   \begin{array}{ll}
   O_1 = \bar q\,\gamma^\mu\,i\,\rlap/\!D\,h_v \,, &\qquad
   O_4 = \bar q\,(-i v\!\cdot\!\overleftarrow{D})\,\gamma^\mu h_v \,,
   \\
   O_2 = \bar q\,v^\mu\,i\,\rlap/\!D\,h_v \,, &\qquad
   O_5 = \bar q\,(-i v\!\cdot\!\overleftarrow{D})\,v^\mu h_v \,,
   \\
   O_3 = \bar q\,i D^\mu h_v \,, &\qquad
   O_6 = \bar q\,(-i\overleftarrow{D^\mu})\,h_v \,.
   \end{array}
\end{equation}
For simplicity, we consider here the limit where the light quark is
massless. Otherwise one would have to include two additional operators
$O_7=m_q\,J_1$ and $O_8=m_q\,J_2$. It is convenient to work with a
regularization scheme with anticommuting $\gamma_5$. This has the
advantage that, to all orders in $1/m_Q$, the operator product
expansion of the axial vector current can be simply obtained from
(\ref{current}) by replacing $\bar q\to-\bar q\,\gamma_5$ in the HQET
operators. The Wilson coefficients remain unchanged. The reason is that
in any diagram the $\gamma_5$ from the current can be moved outside
next to the light quark spinor. For $m_q=0$, this operation always
leads to a minus sign. Hence it is sufficient to consider the case of
the vector current.

A ``hidden'' symmetry of the effective theory, namely its invariance
under reparameterizations of the heavy quark velocity and residual
momentum which leave the total momentum unchanged \cite{LuMa},
determines three of the coefficients $B_i(\mu)$. It implies that, to
all orders in perturbation theory \cite{Matt},
\begin{eqnarray}\label{B123}
   B_1(\mu) &=& C_1(\mu) \,, \nonumber\\
   B_2(\mu) &=& {1\over 2}\,B_3(\mu) = C_2(\mu) \,.
\end{eqnarray}
The remaining coefficients in (\ref{current}) can be obtained from the
solution of the renormalization group equation that determines the
scale dependence of the renormalized current operators in HQET. For our
purposes, it will be sufficient to know these coefficients in leading
logarithmic approximation. They are \cite{Volo,AMM,FG}
\begin{eqnarray}\label{Wilson}
   \phantom{ \bigg[ }
   C_1(\mu) &=& x^{2/\beta} \,, \qquad C_2(\mu) = 0 \,, \nonumber\\
   B_4(\mu) &=& {34\over27}\,x^{2/\beta} - {4\over27}\,x^{-1/\beta}
    - {10\over9} + {16\over3\beta}\,x^{2/\beta}\,\ln x \,, \nonumber\\
   B_5(\mu) &=& - {28\over27}\,x^{2/\beta} + {88\over27}\,x^{-1/\beta}
    - {20\over9} \,, \nonumber\\
   B_6(\mu) &=& - 2\,x^{2/\beta} - {4\over3}\,x^{-1/\beta}
    + {10\over3} \,,
\end{eqnarray}
where again $x=\alpha(\mu)/\alpha(m_Q)$.

After the effective Lagrangian and currents have been constructed, one
proceeds to parameterize the relevant hadronic matrix elements of the
HQET operators in terms of universal, $m_Q$-independent form factors.
In the effective theory, hadrons containing a heavy quark can be
represented by covariant tensor wave functions, which are determined
completely by their transformation properties under the Lorentz group
and heavy quark symmetry. In particular, the ground-state
pseudoscalar and vector mesons are described by \cite{Falk,Bjor}
\begin{equation}
   {\cal{M}}(v) = {1+\rlap/v\over 2}\,
   \cases{ -\gamma_5 &; pseudoscalar meson, \cr
           \rlap/\epsilon &; vector meson. \cr}
\end{equation}
Here $\epsilon^\mu$ is the polarization vector of the vector meson. Any
matrix element of an operator of the effective theory can be written as
a trace over such wave functions, whose structure is determined by
symmetry and by the Feynman rules of the effective theory.

We will now develop this formalism for $B\to\pi$ transitions. Matrix
elements of the leading-order currents $J_i$ in (\ref{LOcurrent}) can
be written as (see, e.g., Ref.~\cite{Thom})
\begin{equation}\label{LOtrace}
  \langle\pi(p)|\,\bar q\,\Gamma\,h_v\,|M(v)\rangle
  = - {\rm Tr}\Big\{ \Pi(v,p)\,\Gamma\,{\cal{M}}(v) \Big\} \,,
\end{equation}
where $\Gamma$ is an arbitrary Dirac matrix. Note that we use a mass
independent normalization of meson states to $2 v^0$ (instead of
$2 p^0$), as this is more convenient when dealing with heavy quark
systems. The Feynman rules of HQET imply that there cannot appear any
$\gamma$-matrices on the right-hand side of $\Gamma$. The matrix
$\Pi(v,p)$ must transform as a pseudoscalar, but is otherwise a general
function of $v$ and $p$. Using the fact that ${\cal{M}}(v)\,\rlap/v =
-{\cal{M}}(v)$, we can write down the most general decomposition
\begin{equation}
  \Pi(v,p) = \gamma_5\,\Big[ A(v\cdot p,\mu)
  + \,\rlap/\!\widehat p\,B(v\cdot p,\mu) \Big] \,.
\end{equation}
We find it convenient to introduce the dimensionless variable
\begin{equation}
   \widehat p^\mu = {p^\mu\over v\cdot p} \,,\qquad
   v\cdot\widehat p = 1 \,,
\end{equation}
so that the scalar functions $A(v\cdot p,\mu)$ and $B(v\cdot p,\mu)$
have the same dimension. These universal form factors depend on the
kinematic variable $v\cdot p$. They also depend on the scale $\mu$ at
which the HQET operators are renormalized, but not on the heavy quark
mass $m_Q$. These functions are the analogs of the celebrated
Isgur-Wise function, which describes heavy-to-heavy meson transitions
at leading order in HQET \cite{Isgu}.

Let us now turn to the study of the leading power corrections
proportional to $1/m_Q$, which arise from the corrections both to the
currents and to the effective Lagrangian of HQET. We first consider the
dimension-four operators in the expansion of the currents
(\ref{current}). Matrix elements of the operators $O_1$, $O_2$, and
$O_3$, which contain a covariant derivative acting on the heavy quark
field, have the generic structure
\begin{eqnarray}\label{Fdef}
   \langle\pi(p)|\,\bar q\,\Gamma\,i D^\mu h_v\,|M(v)\rangle
   = - {\rm Tr}\Big\{ \Big[ &(& F_1\,v^\mu + F_2\,\widehat p^\mu
    + F_3\,\gamma^\mu)\,\gamma_5 \nonumber\\
   + &(& F_4\,v^\mu + F_5\,\widehat p^\mu
    + F_6\,\gamma^\mu)\,\gamma_5\,\,\rlap/\!\widehat p\,\Big]\,
    \Gamma\,{\cal{M}}(v) \Big\} \,.
\end{eqnarray}
The functions $F_i(v\cdot p,\mu)$ are new low-energy parameters. They,
again, depend only on the kinematic variable $v\cdot p$ and the
renormalization scale (although we do not display this dependence for
simplicity), but not on the heavy quark mass. Not all of these
functions are independent. The equation of motion, $i v\!\cdot\!D\,
h_v=0$, implies
\begin{eqnarray}\label{rel12}
   F_1 + F_2 - F_3 &=& 0\,, \nonumber\\
   F_4 + F_5 - F_6 &=& 0\,.
\end{eqnarray}
We may furthermore use the structure of the field redefinition
(\ref{field}) to derive that
\begin{equation}\label{totalder}
   \langle\pi(p)|\,i\partial^\mu(\bar q\,\Gamma\,h_v)\,|M(v)\rangle
   = (\bar\Lambda\,v^\mu - p^\mu)\,
   \langle\pi(p)|\,\bar q\,\Gamma\,h_v\,|M(v)\rangle \,,
\end{equation}
where $\bar\Lambda=m_M-m_Q$ denotes the finite mass difference between
a heavy meson and the heavy quark that it contains, in the infinite
quark mass limit \cite{Luke,AMM}. This parameter sets the canonical
scale for power corrections in HQET. Substituting
$\Gamma=\gamma_\mu\,\Gamma'$ into the above relation, and using the
equation of motion for the light quark field, $i\,\rlap/\!D\,q=0$, we
find
\begin{eqnarray}\label{rel34}
   F_2 - F_4 + 2\,F_6 &=& - v\cdot p\,A - \bar\Lambda\,B \,,
    \nonumber\\
   F_1 - 4\,F_3 + 2\,F_4 + \widehat p^2\,F_5 &=& \bar\Lambda\,A
    + (2\bar\Lambda - v\cdot p\,\widehat p^2)\,B \,.
\end{eqnarray}
We shall use the relations (\ref{rel12}) and (\ref{rel34}) to eliminate
$F_1$, $F_2$, $F_3$, and $F_4$ in favor of $F_5$ and $F_6$. Matrix
elements of the operators $O_4$, $O_5$, and $O_6$ in (\ref{SLOcurrent})
can be evaluated along the same lines, using
\begin{equation}
   \bar q\,(-i\overleftarrow{D^\mu})\,\Gamma\,h_v
   = \bar q\,\Gamma\,(i D^\mu)\,h_v - i\partial^\mu
   (\bar q\,\Gamma\,h_v)
\end{equation}
together with (\ref{totalder}).

Next we investigate the effects of $1/m_Q$ corrections to the effective
Lagrangian of HQET. The operators $O_{\rm kin}$ and $O_{\rm mag}$ in
(\ref{Omag}) can be inserted into matrix elements of the leading-order
currents $J_i$. The corresponding corrections can be described in terms
of six additional functions $G_i(v\cdot p,\mu)$, which parameterize the
matrix elements of the time-ordered products
\begin{eqnarray}\label{Gdef}
   &&\langle\pi(p)|\,i\int{\rm d}y\,T\Big\{ \bar q\,\Gamma\,h_v(0),
   O_{\rm kin}(y) \Big\}\,|M(v)\rangle
   = - {\rm Tr}\Big\{ \gamma_5\,(G_1 + \,\rlap/\!\widehat p\,G_2)
    \,\Gamma\,{\cal{M}}(v) \Big\} \,, \nonumber\\
   && \\
   &&\langle\pi(p)|\,i\int{\rm d}y\,T\Big\{ \bar q\,\Gamma\,h_v(0),
   O_{\rm mag}(y) \Big\}\,|M(v)\rangle \nonumber\\
   &&\quad = - {\rm Tr}\bigg\{ \Big[ (i G_3\,\widehat p_\alpha
    \gamma_\beta + G_4\,\sigma_{\alpha\beta})\,\gamma_5
    + (i G_5\,\widehat p_\alpha\gamma_\beta
    + G_6\,\sigma_{\alpha\beta})\,\gamma_5\,
    \,\rlap/\!\widehat p\,\Big]\,\Gamma\,{1+\rlap/v\over 2}\,
    \sigma^{\alpha\beta}\,{\cal{M}}(v) \bigg\} \,. \nonumber
\end{eqnarray}

Using the above definitions and relations, it is a matter of patience
to compute the matrix elements relevant to $B^{(*)}\to\pi\,\ell\,\nu$
to order $1/m_b$. We will discuss these matrix elements in the
following section.

\section{Matrix elements}
\label{sec:3}

The matrix element of the flavor-changing vector current responsible
for the decay $B\to\pi\,\ell\,\nu$ can be parameterized in terms of
two invariant form factors, which are conveniently defined as
\begin{eqnarray}\label{fpm}
   \sqrt{m_B}\,\langle\pi(p)|\,\bar q\,\gamma^\mu\,Q\,|B(v)\rangle
   &=& f_+(q^2)\,\bigg[ (m_B\,v+p)^\mu - {m_B^2-m_\pi^2\over q^2}\,
    q^\mu \bigg] \nonumber\\
   &+& f_0(q^2)\,{m_B^2-m_\pi^2\over q^2}\,q^\mu \,,
\end{eqnarray}
where $q= m_B\,v-p$. The prefactor $\sqrt{m_B}$ appears since we use a
somewhat unconventional normalization of states. In practice, only
$f_+(q^2)$ is measurable in $B\to\pi\,\ell\,\nu$ decays into the light
leptons $e$ or $\mu$, since the contribution of $f_0(q^2)$ to the decay
rate is suppressed by a factor $m_\ell^2/m_B^2$. However, both form
factors are important in $B\to\pi\,\tau\,\nu$ decays.

As we have seen above, in the context of HQET it is more natural to
work with the velocity of the heavy meson, and to consider the form
factors as functions of the kinematic variable
\begin{equation}
   v\cdot p = {m_B^2 + m_\pi^2 - q^2\over 2 m_B} \,.
\end{equation}
Accordingly, we define
\begin{equation}\label{ffdef}
   \langle\pi(p)|\,\bar q\,\gamma^\mu\,Q\,|B(v)\rangle
   = 2\,\Big[ f_1(v\cdot p)\,v^\mu + f_2(v\cdot p)\,\widehat p^\mu
   \Big] \,.
\end{equation}
The two sets of form factors are related by
\begin{eqnarray}\label{ffrela}
   f_+(q^2) &=& \sqrt{m_B}\,\bigg\{ {f_2(v\cdot p)\over v\cdot p}
    + {f_1(v\cdot p)\over m_B} \bigg\} \,, \nonumber\\
   f_0(q^2) &=& {2\over\sqrt{m_B}}\,{m_B^2\over m_B^2 - m_\pi^2}\,
    \bigg\{ \Big[ f_1(v\cdot p) + f_2(v\cdot p) \Big]
    - {v\cdot p\over m_B}\,\Big[ f_1(v\cdot p) + \widehat p^2\,
    f_2(v\cdot p) \Big] \bigg\} \,.
\end{eqnarray}
The fact that in the $m_b\to\infty$ limit the functions $f_{1,2}
(v\cdot p)$ become independent of $m_b$ (modulo logarithms) implies the
well-known scaling relations \cite{Vub}
\begin{equation}
   f_+ \sim \sqrt{m_B} \,, \qquad
   f_0 \sim 1/\sqrt{m_B} \,,
\end{equation}
which are valid as long as $v\cdot p$ does not scale with $m_B$.

By evaluating the traces and using the definitions of the previous
section, we find, to next-to-leading order in $1/m_b$, the following
expressions:
\begin{eqnarray}\label{horrible}
   f_1 &=& C_1\,A + C_2\,(A + B) \nonumber\\
   &+& {1\over 2 m_b}\,\bigg\{
    C_1\,\Big[ - (\bar\Lambda - 2 v\cdot p)\,A
    + v\cdot p\,\widehat p^2\,B + 4 F_6 + G_1 \Big] \nonumber\\
   &&\qquad \mbox{}+ C_2\,\Big[ (\bar\Lambda + v\cdot p)\,A
    + (3\bar\Lambda - v\cdot p\,\widehat p^2)\,B + 4 F_6
    + G_1 + G_2 \Big] \nonumber\\
   &&\qquad \mbox{}- B_4\,(\bar\Lambda - v\cdot p)\,A
    - B_5\,(\bar\Lambda - v\cdot p)\,(A+B) - B_6\,\Big[
    (\bar\Lambda - v\cdot p)\,A - 2 F_6 \Big] \nonumber\\
   &&\qquad \mbox{}+ C_1\,C_{\rm mag}\,\Big[ - 2 G_3 + 6 G_4
    + 2\widehat p^2\,G_5 \Big] + C_2\,C_{\rm mag}\,
    \Big[ 6 G_4 - 2(1-\widehat p^2)\,G_5 + 6 G_6 \Big] \bigg\}
    \,, \nonumber\\
   && \\
   f_2 &=& C_1\,B + {1\over 2 m_b}\,\bigg\{ C_1\,
    \Big[ - v\cdot p\,A - \bar\Lambda\,B - 4 F_6 + G_2\Big]
    - 2 C_2\,\Big[ v\cdot p\,A + \bar\Lambda\,B + 2 F_6 \Big]
    \nonumber\\
   &&\qquad \mbox{}- B_4\,(\bar\Lambda - v\cdot p)\,B
    - B_6\,\Big[ (\bar\Lambda - v\cdot p)\,B + 2 F_6 \Big] \nonumber\\
   &&\qquad \mbox{}+ C_1\,C_{\rm mag}\,\Big[ 2 G_3 - 2 G_5 + 6 G_6
    \Big] \bigg\} \,. \nonumber
\end{eqnarray}
For simplicity, we have omitted the dependence of the universal
functions on $v\cdot p$ and $\mu$, and the dependence of the Wilson
coefficients on $\mu$.

{}From the fact that the physical form factors must be independent of the
renormalization scale, one can deduce the $\mu$-dependence of the
universal functions of HQET, since it has to cancel against that of the
Wilson coefficients. For the leading-order functions $A(v\cdot p,\mu)$
and $B(v\cdot p,\mu)$, it follows that
\begin{eqnarray}
   A_{\rm ren}(v\cdot p) &\equiv&
    \Big[ \alpha_s(\mu) \Big]^{2/\beta}\,A(v\cdot p,\mu) \,,
    \nonumber\\
   B_{\rm ren}(v\cdot p) &\equiv&
    \Big[ \alpha_s(\mu) \Big]^{2/\beta}\,B(v\cdot p,\mu)
\end{eqnarray}
must be $\mu$-independent (in leading logarithmic approximation). It is
then convenient to define two related functions
\begin{eqnarray}\label{Ahatdef}
   \hat A(v\cdot p) &\equiv&
    \Big[ \alpha_s(m_Q) \Big]^{-2/\beta}\,A_{\rm ren}(v\cdot p)
    = x^{2/\beta}\,A(v\cdot p,\mu) \,, \nonumber\\
   \hat B(v\cdot p) &\equiv&
    \Big[ \alpha_s(m_Q) \Big]^{-2/\beta}\,B_{\rm ren}(v\cdot p)
    = x^{2/\beta}\,B(v\cdot p,\mu) \,,
\end{eqnarray}
which are clearly also $\mu$-independent. These functions are no longer
universal since they contain a logarithmic dependence on the heavy
quark mass. At tree-level, however, they agree with the original
functions $A$ and $B$.

In order to find the corresponding relations for the subleading
universal functions $F_i$ and $G_i$, the expressions (\ref{horrible})
for $f_1$ and $f_2$ are not sufficient. We have thus worked out the
heavy quark expansion for $B^*\to\pi\,\ell\,\nu$ decays, although these
processes have little (if any) phenomenological relevance. We define
hadronic form factors $h_i$ by
\begin{eqnarray}\label{hidef}
   \langle\pi(p)|\,\bar q\,\gamma^\mu\,Q\,|B^*(v)\rangle
   &=& 2 i\,\varepsilon^{\mu\alpha\beta\gamma}\,\epsilon_\alpha\,
    \widehat p_\beta\,v_\gamma\,h_1(v\cdot p) \,, \nonumber\\
   && \\
   \langle\pi(p)|\,\bar q\,\gamma^\mu\gamma_5\,Q\,|B^*(v)\rangle
   &=& 2\,\Big[ h_2(v\cdot p)\,\epsilon^\mu
    - h_3(v\cdot p)\,\epsilon\!\cdot\!\widehat p\,\,v^\mu
    - h_4(v\cdot p)\,\epsilon\!\cdot\!\widehat p\,\,\widehat p^\mu
    \Big] \,. \nonumber
\end{eqnarray}
By studying these form factors at order $1/m_b$, one can derive enough
relations to fully determine the $\mu$-dependence of the universal
functions. We discuss this somewhat technical issue in
appendix~\ref{app:1}. There we define a set of renormalization-group
invariant functions $\hat F_i(v\cdot p)$ and $\hat G_i(v\cdot p)$,
which are $\mu$-independent and coincide with $F_i$ and $G_i$ at
tree-level. In terms of these functions, the form factor relations take
a much simpler form. Instead of (\ref{horrible}), we find
\begin{eqnarray}\label{mtrxel}
   f_1 &=& \hat A + {1\over 2 m_b}\,\Big[ -(\bar\Lambda-2 v\cdot p)\,
    \hat A + v\cdot p\,\widehat p^2\,\hat B + 4\hat F_6 + \hat G_1
    -2\hat G_3 + 6\hat G_4 + 2\widehat p^2\,\hat G_5 \Big] \,,
    \nonumber\\
   f_2 &=& \hat B + {1\over 2 m_b}\,\Big[ -v\cdot p\,\hat A
    - \bar\Lambda\,\hat B - 4\hat F_6 + \hat G_2 + 2\hat G_3
    - 2\hat G_5 + 6\hat G_6 \Big] \,,
\end{eqnarray}
and the $B^*\to\pi$ decay form factors are given by
\begin{eqnarray}
   h_1 &=& \hat B + {1\over 2 m_b}\,\Big[ v\cdot p\,\hat A
    + \bar\Lambda\hat B + 2\hat F_6 + \hat G_2 - 2\hat G_6 \Big]
    \,, \nonumber\\
   h_2 &=& (\hat A + \hat B) + {1\over 2 m_b}\,\bigg[ \,{1\over 3}\,
    (\bar\Lambda-v\cdot p)\,\hat A + {1\over 3}\,
    (\bar\Lambda-v\cdot p\,\widehat p^2)\,\hat B - {2\over 3}\,
    (1-\widehat p^2)\,\hat F_5 \nonumber\\
   &&\phantom{ (\hat A + \hat B) + {1\over 2 m_b}\,\bigg[ }
    \mbox{}+ \hat G_1 + \hat G_2 - 2\hat G_4 - 2\hat G_6 \bigg] \,,
     \nonumber\\
   h_3 &=& \hat B + {1\over 2 m_b}\,\Big[ -v\cdot p\,\hat A
    - \bar\Lambda\,\hat B - 2\hat F_5 + \hat G_2 - 2\hat G_3
    - 2\hat G_6 \Big] \,, \nonumber\\
   h_4 &=& {1\over 2 m_b}\,\Big[ 2\hat F_5 + 2\hat G_5 \Big] \,.
\end{eqnarray}
Note that $\hat F_6$ appears only in the vector form factors $f_1$,
$f_2$, and $h_1$, whereas $\hat F_5$ appears only in the axial form
factors $h_2$, $h_3$, and $h_4$.

Most relevant, of course, are the form factors $f_+$ and $f_0$ that are
usually used to describe $B\to\pi\,\ell\,\nu$ decays. From
(\ref{ffrela}),  we obtain at next-to-leading order in $1/m_b$:
\begin{eqnarray}\label{fplusf0}
   f_+ &=& {\sqrt{m_B}\over v\cdot p}\,\bigg\{ \hat B
    + {1\over 2 m_b}\,\Big[ v\cdot p\,\hat A - \bar\Lambda\,\hat B
    - 4\hat F_6 + \hat G_2 + 2\hat G_3 - 2\hat G_5 + 6\hat G_6 \Big]
    \bigg\} \,, \nonumber\\
   && \\
   f_0 &=& {2\over\sqrt{m_B}}\,\bigg\{ (\hat A + \hat B)
    + {1\over 2 m_b}\,\Big[ - (\bar\Lambda + v\cdot p)\,\hat A
    - (\bar\Lambda + v\cdot p\,\widehat p^2)\,\hat B \nonumber\\
   &&\phantom{ {2\over\sqrt{m_B}}\,\bigg\{ (\hat A + \hat B)
               + {1\over 2 m_b}\,\Big[ }
    \mbox{}+ \hat G_1 + \hat G_2 + 6\hat G_4 - 2(1-\widehat p^2)
    \hat G_5 + 6\hat G_6 \Big] \bigg\} \,. \nonumber
\end{eqnarray}
These relations show how, in a rather complicated way, the $1/m_b$
corrections to $f_+$ and $f_0$ are related to matrix elements of
operators in HQET. To gain more insight into the structure of the
corrections, it is instructive to consider the soft pion limit
$v\cdot p\to 0$ and $p^2=m_\pi^2\to 0$, in which current algebra can be
used to derive normalization conditions on the universal functions.
This is the subject of the following section.

\section{Soft pion relations}
\label{sec:4}

In this section we shall derive the normalization conditions for the
universal functions of HQET, which arise in the soft pion limit
$p\to 0$. Our goal is to reduce, as much as possible, the number of
independent parameters upon which our predictions depend. The soft pion
relations are derived by using the PCAC relation for the pion field. To
be specific, let us consider the decay $M^0\to\pi^+\ell^-\nu$
(where $M^0=\bar B^0$ or $\bar B^{0*}$). Then
\begin{equation}
   \pi^+(x) = {1\over f_\pi m_\pi^2}\,\partial^\mu A_\mu(x) \,,
\end{equation}
where $A_\mu=\bar d\,\gamma_\mu\gamma_5\,u$ is the axial vector
current, and $f_\pi\simeq 132$ MeV is the pion decay constant. The LSZ
reduction formalism can be employed to write
\begin{equation}\label{soft1}
   \langle\pi(p)|\,O(0)\,|M(v)\rangle = \lim_{p^2\to m_\pi^2}\,
   {1\over f_\pi}\,{m_\pi^2-p^2\over m_\pi^2}\,\,
   i\int{\rm d}x\,e^{ip\cdot x}\,\langle\,0\,|\,T\,\big\{
   O(0),\partial^\mu A_\mu(x) \big\}\,|M(v)\rangle \,,
\end{equation}
where $O$ may be any operator that couples $M$ to $\pi$. The right-hand
side can be rewritten using
\begin{equation}\label{ident}\label{soft2}
   i\int{\rm d}x\,e^{ip\cdot x}\,\,T\,\big\{ O(0),\partial^\mu
   A_\mu(x) \big\} = p^\mu \int{\rm d}x\,e^{ip\cdot x}\,\,T\,
   \big\{ A_\mu(x),O(0) \big\} -i\,\big[ Q_5,O(0) \big] \,.
\end{equation}
Here $Q_5$ denotes the axial charge, i.e., the spatial integral of
the zero component of $A_\mu$: $Q_5=\int{\rm d}^3x\,d^\dagger
\gamma_5\,u$. Therefore,
\begin{equation}
   \big[ Q_5,O(0) \big] = O'(0) \,,
\end{equation}
where the operator $O'$ is obtained from $O$ by replacing $\bar u$ by
$\bar d(-\gamma_5)$, i.e., if $O=\bar u\,\gamma^\mu\,b$ then $O'=
\bar d\,\gamma^\mu\gamma_5\,b$, etc. The soft pion relation is obtained
by analytically continuing (\ref{soft1}) to $p\to 0$. In this limit the
first term on the right-hand side of (\ref{soft2}) is saturated by
intermediate states degenerate with the ground-state. They lead to
poles proportional to $1/v\cdot p$, which cancel the factor $p^\mu$ in
front of the integral. In the case of $B\to\pi$ transitions, the
relevant intermediate state will be the $B^*$ meson, which to leading
order in HQET is in fact degenerate with the $B$ meson. We obtain
\begin{eqnarray}\label{softpi}
   &&\lim_{p\to 0}\,\langle\pi(p)|\,O(0)\,|M(v)\rangle \nonumber\\
   &&\quad = {1\over f_\pi}\,\bigg\{
    -i\,\langle\,0\,|\,O'(0)\,|M(v)\rangle + \lim_{p\to0}\,
    \int{\rm d}x\,e^{ip\cdot x}\,\langle\,0\,|\,T\,\big\{
    O(0),p\cdot A(x) \big\}\,|M(v)\rangle \bigg\} \,.
\end{eqnarray}
In what follows we shall refer to the first and second terms on the
right-hand side as the commutator and the pole contribution,
respectively.

\subsection{Soft pion relations for $\hat A(v\cdot p)$ and
$\hat B(v\cdot p)$}

Let us now evaluate this relation for the matrix elements arising at
leading order in the $1/m_Q$ expansion, where the effective current
operators have the generic form $O=\bar q\,\Gamma\,h_v$. Both the
commutator and the pole contribution involve a current-induced
transition of a heavy meson into the vacuum. At leading order in HQET,
the corresponding matrix elements can be written as \cite{MN}
\begin{equation}
   \langle\,0\,|\,\bar q\,\Gamma'\,h_v\,|M(v)\rangle
   = {i F(\mu)\over2}\,{\rm Tr}\big\{ \Gamma' {\cal{M}}(v) \big\}
   \,,
\end{equation}
where $\Gamma'=-\gamma_5\,\Gamma$ in the commutator term, and
$\Gamma'=\Gamma$ in the pole term. The prefactor is chosen such that
the universal low-energy parameter $F(\mu)$, which is independent of
the heavy quark mass, corresponds to the asymptotic value of the scaled
meson decay constant: $F\sim f_M\sqrt{m_M}$ (modulo logarithms).

To compute the pole term, we further need the coupling of two heavy
mesons to the axial vector current, as shown in Fig.~\ref{fig:1}. We
define
\begin{equation}
   \langle M'(v,p)|\,p\cdot A\,|M(v,0)\rangle
   = g(v\cdot p)\,{\rm Tr}\Big\{ \gamma_5\,\,\rlap/\!p\,
   \overline{\cal{M}}'(v){\cal{M}}(v) \Big\} \,,
\end{equation}
where $M'$ is off-shell by the pion momentum $p$. The form factor
$g(v\cdot p)$ is real and regular as $v\cdot p\to 0$. We define
\begin{equation}\label{gpidef}
   \lim_{p\to 0}\,g(v\cdot p) = g(0) \equiv g \,.
\end{equation}
Note that $g$ is renormalization-group invariant. We can now write the
pole contribution as
\begin{eqnarray}\label{LOpole}
   &&\sum_{M'}\,\langle\,0\,|\,\bar q\,\Gamma\,h_v\,|M'(v)\rangle\,
    {i\over 2 v\cdot(-p)}\,\langle M'(v)|\,p\cdot A\,|M(v)\rangle
    \nonumber\\
   &&\quad = {F(\mu)\,g(v\cdot p)\over 4 v\cdot p}\,\sum_{M'}\,
    {\rm Tr}\big\{ \Gamma\,{\cal{M}}'(v) \big\}\,
    {\rm Tr}\Big\{ \gamma_5\,\,\rlap/\!p\,\overline{\cal{M}}'(v)
    {\cal{M}}(v) \Big\} \,,
\end{eqnarray}
where we have used that in the effective theory the intermediate meson
propagator is simply given by $i/v\cdot k$, where $k$ stands for the
residual momentum. (Recall that we use a mass independent normalization
of states.) To proceed further, we need a relation that allows us to
combine the two traces appearing on the right-hand side into a single
trace. This is accomplished by the identity
\begin{equation}\label{magictrace}
   \sum_{M'=P,V} {\rm Tr}\big\{ X\,{\cal{M'}}(v) \big\}\,
   {\rm Tr}\Big\{ \gamma_5\,\,\rlap/\!p\,\overline{\cal{M}}'(v)
   {\cal{M}}(v) \Big\} = -2\,{\rm Tr}\Big\{ \gamma_5\,
   (\,\rlap/\!p - v\cdot p)\,X\,{\cal{M}}(v) \Big\} \,,
\end{equation}
which is valid for any Dirac matrix $X$, and for a pseudoscalar or
vector meson $M(v)$. The sum extends over the ground-state pseudoscalar
and vector mesons $M'(v)$ degenerate with $M(v)$, and summation over
polarizations is understood if $M'$ is a vector meson.

Putting together the various pieces and using (\ref{LOtrace}), we
obtain the soft pion relation
\begin{eqnarray}
   &&\lim_{p\to 0}\,{\rm Tr}\Big\{ \gamma_5\,\Big[ A(v\cdot p,\mu)
    + \,\rlap/\!\widehat p\,B(v\cdot p,\mu) \Big]\,\Gamma\,
    {\cal{M}}(v) \Big\} \nonumber\\
   &&\quad = {F(\mu) \over 2 f_\pi}\,\bigg[
    \,{\rm Tr}\big\{ \gamma_5\,\Gamma\,{\cal{M}}(v) \big\}
    + \lim_{p\to0}\,g(v\cdot p)\,{\rm Tr}\Big\{ \gamma_5\,
    (\,\rlap/\!\widehat p-1)\,\Gamma\,{\cal{M}}(v) \Big\}  \bigg] \,,
\end{eqnarray}
from which we read off the values of the form factors $A$ and $B$ in
the soft pion limit:
\begin{equation}
   A(0,\mu) = {F(\mu)\over 2 f_\pi}\,(1 - g) \,, \qquad
   B(0,\mu) = {F(\mu)\over 2 f_\pi}\,g \,.
\end{equation}
These relations are preserved by renormalization. In fact, one can
define a scale-independent quantity $\hat F=x^{2/\beta}\,F(\mu)$, which
agrees with $F$ at tree-level \cite{MN}. As mentioned above, the
coupling constant $g$ is not renormalized. From (\ref{Ahatdef}), it
then follows that
\begin{equation}\label{softAB}
   \hat A(0) = {\hat F\over 2 f_\pi}\,(1 - g) \,, \qquad
   \hat B(0) = {\hat F\over 2 f_\pi}\,g \,,
\end{equation}
which are the desired normalization conditions for the universal
functions in the soft pion limit.

\subsection{Soft pion relations for $\hat F_i(v\cdot p)$}

Let us next consider the soft pion relations for the subleading form
factors $F_i$ defined in (\ref{Fdef}). The only difference from the
previous derivation is that now the current contains a covariant
derivative. Hence we need the corresponding matrix elements for the
case of meson-to-vacuum transitions. They are \cite{MN}
\begin{equation}\label{Flamtr}
   \langle\,0\,|\,\bar q\,\Gamma'\,i D^\mu h_v\,|M(v)\rangle
   = - {i F(\mu)\over 2}\,{\bar\Lambda\over 3}\,{\rm Tr}\Big\{
   (v^\mu + \gamma^\mu)\,\Gamma'{\cal{M}}(v) \Big\} \,.
\end{equation}
Using again the trace relation (\ref{magictrace}), we obtain for the
pole term:
\begin{eqnarray}
   && \sum_{M'}\,\langle\,0\,|\,\bar q\,\Gamma\,i D^\mu h_v\,
    |M'(v)\rangle\,{i\over 2 v\cdot(-p)}\,\langle M'(v)|\,p\cdot A\,
    |M(v)\rangle \nonumber\\
   &&\quad = {F(\mu)\,\bar\Lambda\over 6}\,{g(v\cdot p)\over v\cdot p}
    \,{\rm Tr}\Big\{ \gamma_5\,(\,\rlap/\!p - v\cdot p)\,
    (v^\mu + \gamma^\mu)\,\Gamma\,{\cal{M}}(v) \Big\} \,.
\end{eqnarray}
The commutator term is simply given by (\ref{Flamtr}) with
$\Gamma'=-\gamma_5\,\Gamma$. Combining the two, we obtain from a
comparison with (\ref{Fdef}):
\begin{equation}
   \begin{array}{ll}
\displaystyle
   F_1(0,\mu) = - {F(\mu)\over 2 f_\pi}\,{\bar\Lambda\over 3}\,(1-g)
    \,, &\qquad
\displaystyle
   F_4(0,\mu) = - {F(\mu)\over 2 f_\pi}\,{\bar\Lambda\over 3}\,g \,,
   \\ [12pt]
\displaystyle
   F_2(0,\mu) = - {F(\mu)\over 2 f_\pi}\,{2\bar\Lambda\over 3}\,g
    \,, &\qquad
\displaystyle
   F_5(0,\mu) = 0 \,,
   \\ [12pt]
\displaystyle
   F_3(0,\mu) = - {F(\mu)\over 2 f_\pi}\,{\bar\Lambda\over 3}\,(1+g)
    \,, &\qquad
\displaystyle
   F_6(0,\mu) = - {F(\mu)\over 2 f_\pi}\,{\bar\Lambda\over 3}\,g \,.
   \end{array}
\end{equation}
Note that the relations (\ref{rel12}) and (\ref{rel34}), which are
consequences of the equations of motion, are satisfied by these
expressions. Using the results of appendix~\ref{app:1}, we find that
radiative corrections can again be incorporated in a straightforward
manner. The two independent renormalized form factors satisfy
\begin{equation}\label{softF}
   \hat F_5(0) = 0 \,, \qquad
   \hat F_6(0) = - {\hat F\over 2 f_\pi}\,{\bar\Lambda\over 3}\,g \,.
\end{equation}

\subsection{Soft pion relations for $\hat G_i(v\cdot p)$}

Here one encounters the complication that the soft pion relation
involves the time-ordered product of three operators: the original
heavy-light current, the axial vector current that interpolates the
pion field, and one of the operators $O_{\rm kin}$ and $O_{\rm mag}$
which appear at order $1/m_Q$ in the effective Lagrangian of HQET.
Consequently, there are both single and double pole contributions in
addition to the commutator term, and the derivations become more
cumbersome. We shall only give the final expressions here and refer the
interested reader to appendix~\ref{app:2}, where we give details of the
calculation.

The $1/m_Q$ insertions from corrections to the effective Lagrangian
correct both the meson decay constants and the $M M'\pi$ coupling
constant, as shown in Fig.~\ref{fig:2}a. The corrections to the decay
constant were treated in Ref.~\cite{MN}. They can be parameterized in
terms of two renormalized parameters $\hat{\cal{G}}_1$ and
$\hat{\cal{G}}_2$, which describe the effects of the kinetic and
chromo-magnetic operator, respectively. To order $1/m_Q$ (and in
leading logarithmic approximation), the physical decay constants are
\begin{eqnarray}\label{fdecay}
   f_P\sqrt{m_P} &=& \hat F\,\bigg\{ 1 + {1\over m_Q}\,\bigg(
    \hat{\cal{G}}_1 + 6\hat{\cal{G}}_2 - {\bar\Lambda\over 2} \bigg)
    \bigg\} \,, \nonumber\\
   f_V\sqrt{m_V} &=& \hat F\,\bigg\{ 1 + {1\over m_Q}\,\bigg(
    \hat{\cal{G}}_1 - 2\hat{\cal{G}}_2 + {\bar\Lambda\over 6} \bigg)
    \bigg\} \,.
\end{eqnarray}
Similarly, at next-to-leading order in $1/m_Q$ the coupling of two heavy
mesons to the pion receives corrections. Instead of the universal
coupling constant $g$ in (\ref{gpidef}) we write
\begin{eqnarray}\label{gphys}
   g_{PV\pi} = g_{VP\pi} &=& g + {1\over 2 m_Q}\,
    \big( g_1 + 4\hat g_2 \big) \,, \nonumber\\
   g_{VV\pi} &=& g + {1\over 2 m_Q}\,\big( g_1 - 4\hat g_2 \big) \,,
\end{eqnarray}
where $P$ and $V$ stand for a pseudoscalar or vector meson,
respectively. The coupling of two pseudoscalar mesons to the pion
vanishes by parity invariance of the strong interactions. For the
precise definition of the parameters ${\cal{G}}_i$ and $  g_i$ and
their renormalization the reader is encouraged to consult
appendix~\ref{app:2}.

In terms of these parameters, we find the following soft pion relations
for the renormalized form factors $\hat G_i$:
\begin{eqnarray}\label{softG}
   \hat G_1(0) &=& {\hat F\over 2 f_\pi}\,\Big[
    2(1-g)\,\hat{\cal{G}}_1 - g_1 \Big] \,, \nonumber\\
   \hat G_2(0) &=& {\hat F\over 2 f_\pi}\,\Big[
    2 g\,\hat{\cal{G}}_1 + g_1 \Big] \,, \nonumber\\
   \hat G_4(0) &=& {\hat F\over 2 f_\pi}\,\Big[
    2(1-g)\,\hat{\cal{G}}_2 - 2 \hat g_2 \Big] \,, \nonumber\\
   \hat G_5(0) &=& 0 \,, \nonumber\\
   \hat G_6(0) &=& {\hat F\over 2 f_\pi}\,\Big[
    2 g\,\hat{\cal{G}}_2 + 2 \hat g_2 \Big] \,,
\end{eqnarray}
as well as
\begin{equation}\label{softG3}
   \lim_{p\to 0}\,\hat G_3(v\cdot p) = - {\hat F\over 2 f_\pi}\,
   \bigg[ (m_V^2 - m_P^2)\,{g\over 2 v\cdot p} + 8 g\,\hat{\cal{G}}_2
   + 4 \hat g_2 \bigg] \,.
\end{equation}
It might seem surprising that $\hat G_3$ develops a pole as
$v\cdot p\to 0$, with a residue proportional to the mass splitting
between vector and pseudoscalar mesons. However, as we shall see below,
this is exactly what is required to recover the correct pole
contributions predicted by chiral symmetry. The singular behavior of
$\hat G_3$ results from the diagrams depicted in Fig.~\ref{fig:2}b. For
later purposes, we define a regular function $\hat G_3^{\rm reg}
(v\cdot p)$ by
\begin{eqnarray}\label{G3reg}
   \hat G_3^{\rm reg}(v\cdot p) &\equiv& \hat G_3(v\cdot p)
    + {\hat F\,g\over 4 f_\pi}\,{(m_V^2 - m_P^2)\over v\cdot p} \,,
    \nonumber\\
   \hat G_3^{\rm reg}(0) &=& - {\hat F\over 2 f_\pi}\,
   \Big[ 8 g\,\hat{\cal{G}}_2 + 4 \hat g_2 \Big] \,.
\end{eqnarray}

\subsection{Meson form factors in the chiral limit}

The soft pion relations derived above will become more transparent when
we consider the physical meson form factors $f_i$ and $h_i$ defined in
(\ref{ffdef}) and (\ref{hidef}). We start by considering the sum
$f_1+f_2$. In the soft pion limit, we obtain
\begin{equation}
   f_1(0) + f_2(0) = {\hat F\over 2 f_\pi}\,\bigg\{ 1
   + {1\over m_b}\,\bigg( \hat{\cal{G}}_1 + 6\hat{\cal{G}}_2
   - {\bar\Lambda\over 2} \bigg) \bigg\}
   = {f_B\sqrt{m_B}\over 2 f_\pi} \,,
\end{equation}
where we have used (\ref{fdecay}) to write the result in terms of the
physical decay constant $f_B$. Next, consider the form factor $f_2$. We
find
\begin{equation}
   \lim_{p\to 0}\,f_2(v\cdot p) = {\hat F\over 2 f_\pi}\,
   \bigg\{ 1 + {1\over m_b}\,\bigg( \hat{\cal{G}}_1 - 2\hat{\cal{G}}_2
   + {\bar\Lambda\over 6} \bigg) \bigg\}\,
   \bigg\{ g + {1\over 2 m_b}\,\big( g_1 + 4 \hat g_2 \big)
   \bigg\}\,\bigg( 1 - {\Delta_B\over v\cdot p} \bigg) \,,
\end{equation}
where
\begin{equation}
   \Delta_B = {m_{B^*}^2 - m_B^2\over 2 m_b} \approx m_{B^*} - m_B \,,
\end{equation}
and we have factorized various terms in an educated way, so that it is
immediate to identify the decay constant of the $B^*$ meson and the
$B B^*\pi$ coupling constant. In fact, using (\ref{fdecay}) and
(\ref{gphys}) we can rewrite the result as
\begin{equation}
   \lim_{p\to 0}\,f_2(v\cdot p)
   = {f_{B^*}\sqrt{m_{B^*}}\over 2 f_\pi}\,g_{B B^*\pi}\,
   {v\cdot p\over(v\cdot p + \Delta_B)} \,.
\end{equation}
The resummation of the $B^*$-pole term, which is allowed to the order
we are working, has removed the spurious singularity at $v\cdot p=0$,
and we have recovered the physical pole position at $v\cdot
p=-\Delta_B$, corresponding to $q^2=m_{B^*}^2$. This becomes apparent
when we use (\ref{ffrela}) to convert to the conventional form factors
$f_+(q^2)$ and $f_0(q^2)$. In the soft pion limit, they become
\begin{eqnarray}\label{wellknown}
   \lim_{q^2\to m_B^2} f_+(q^2) &=& {m_B\over 2}\,{f_{B^*}\over f_\pi}\,
    {g_{B B^*\pi}\over(v\cdot p + \Delta_B)}
    = {f_{B^*}\over f_\pi}\,{g_{B B^*\pi}\over\big[
    1 - q^2/m_{B^*}^2 \big]} \,, \nonumber\\
   f_0(m_B^2) &=& {f_B\over f_\pi} \,,
\end{eqnarray}
where we have used that $m_{B^*}/m_B=1+{\cal{O}}(1/m_b^2)$. These are
the well-known results for the meson form factors in the chiral limit.
They have been previously derived in the $m_b\to\infty$ limit by
combining HQET with chiral perturbation theory \cite{Wise,Burd}, or by
using current algebra in connection with the fact that the $B$ and
$B^*$ mesons are degenerate to leading order in $1/m_b$ \cite{LW}. The
same relations have also been obtained without a heavy quark expansion,
by assuming nearest pole dominance \cite{IWis,Korn}. We emphasize,
however, that here we have not only recovered these results from a
rigorous expansion in QCD, but we have proven them to hold even at
next-to-leading order in $1/m_b$, and including short-distance
corrections. We find that there are no such corrections to the soft
pion relations once one uses the physical values of the meson decay
constants and of the $B B^*\pi$ coupling constant, as compared to their
values in the $m_b\to\infty$ limit.

In a similar manner, one can derive the soft pion limit for the
$B^*\to\pi$ decay form factors $h_i$ defined in (\ref{hidef}). We obtain
\begin{eqnarray}
   h_1(0) &=& {f_{B^*}\sqrt{m_{B^*}}\over 2 f_\pi}\,g_{B^* B^*\pi} \,,
    \nonumber\\
   h_2(0) &=& {f_{B^*}\sqrt{m_{B^*}}\over 2 f_\pi} \,, \nonumber\\
   h_3(0) &=& {f_B\sqrt{m_B}\over 2 f_\pi}\,g_{B^* B\pi}\,
    {v\cdot p\over(v\cdot p - \Delta_B)} \,, \nonumber\\
   h_4(0) &=& 0 \,.
\end{eqnarray}
Again, at leading order in $1/m_b$ these relations could also be
derived using heavy meson chiral perturbation theory.

\section{Summary and Conclusions}
\label{sec:5}

We have presented a systematic analysis of the $B^{(*)}\to\pi\,\ell\,
\nu$ decay form factors to order $1/m_b$ in the heavy quark expansion,
including a detailed treatment of short-distance corrections. Similar
analyses have been carried out in the past for the semileptonic decays
$B\to D^{(*)}\ell\,\nu$ \cite{Luke} and $\Lambda_b\to\Lambda_c\,\ell\,
\nu$ \cite{GGW}, and for heavy meson decay constants \cite{MN}. As in
these cases, the analysis of the form factors in the context of a heavy
quark expansion provides the theoretical framework for a comprehensive
investigation of the hadronic physics encoded in the universal
functions of HQET, using nonperturbative techniques such as lattice
gauge theory or QCD sum rules. For the decays between two heavy mesons,
this strategy has been very successful and has led to much insight into
the properties of these nonperturbative objects. In particular,
analytic (two-loop) predictions have been obtained for the leading and
subleading Isgur-Wise functions using QCD sum rules \cite{2loop,ZMY},
and first results for the leading-order Isgur-Wise function are
available from lattice gauge theory \cite{LGT1,LGT2}. Previous
predictions for the $B\to\pi\,\ell\,\nu$ form factors, on the other
hand, were obtained using quark models \cite{BSW,ISGW}, or QCD sum
rules in the full theory \cite{SR1,SR2,SR3,SR4,SR5,SR6}. The next step
should be a more detailed analysis in the context of the heavy quark
expansion. Recently, calculations incorporating ingredients of heavy
quark symmetry were performed in the $m_b\to\infty$ limit
\cite{HQS1,HQS2,Xu}. One of the purposes of our paper is to allow an
extension of this type of calculations to order $1/m_b$.

The main motivation for a study of exclusive heavy-to-light decays is
to extract the element $|V_{ub}|$ of the quark mixing matrix in a
reliable, model independent way. The idea is to compare the lepton
spectra in the decays $B\to\pi\,\ell\,\nu$ and $D\to\pi\,\ell\,\nu$,
which are related to each other by heavy quark flavor symmetry
\cite{Vub}. In the limit of vanishing lepton mass, the differential
decay rate is determined by the form factor $f_+$ defined in
(\ref{fpm}):
\begin{equation}
   {{\rm d}\Gamma(B\to\pi\,\ell\,\nu)\over{\rm d}(v\cdot p)}
   = {G_F^2\,m_B\over 12\pi^3}\,|V_{ub}|^2\,
   \Big[ (v\cdot p)^2 - m_\pi^2 \Big]^{3/2}\,|\,f_+\,|^2 \,.
\end{equation}
Hence, the ratio of the two distributions at the same value of
$v\cdot p$ is
\begin{equation}
   {{\rm d}\Gamma(B\to\pi\,\ell\,\nu)/{\rm d}(v\cdot p)\over
    {\rm d}\Gamma(D\to\pi\,\ell\,\nu)/{\rm d}(v\cdot p)}
   \Bigg|_{{\rm same}~v\cdot p} = \bigg| {V_{ub}\over V_{cd}}
   \bigg|^2\,\bigg( {m_B\over m_D} \bigg)^2\,\bigg|
   {\sqrt{m_D}\,f_+^{B\to\pi}\over\sqrt{m_B}\,f_+^{D\to\pi}}
   \bigg|^2 \,.
\end{equation}
In the limit of an exact heavy quark flavor symmetry, the last factor
on the right-hand side equals unity. It is convenient to rewrite this
factor as
\begin{equation}\label{RBDdef}
   {\sqrt{m_D}\,f_+^{B\to\pi}\over\sqrt{m_B}\,f_+^{D\to\pi}}
   \equiv {v\cdot p + \Delta_D\over v\cdot p + \Delta_B}\,
   R_{BD}(v\cdot p) \,,
\end{equation}
where $\Delta_B=m_{B^*}-m_B\simeq 0.05$ GeV and $\Delta_D=m_{D^*}-m_D
\simeq 0.14$ GeV. This definition of $R_{BD}$ takes into account the
dominant momentum dependence for low momenta, which comes from the
presence of the nearby vector meson pole. The difference in the pole
positions for $B\to\pi\,\ell\,\nu$ and $D\to\pi\,\ell\,\nu$ is formally
of order $1/m_Q$, but is significant for $v\cdot p$ close to its
minimum value $m_\pi$. This effect is explicitly taken into account in
(\ref{RBDdef}). The remaining, nontrivial power corrections reside in
the quantity $R_{BD}$. Using (\ref{fplusf0}), we obtain
\begin{equation}\label{RBD}
   R_{BD}(v\cdot p) = 1 + {\bar\Lambda\over 2 m_c}\,r_c(v\cdot p)
   - {\bar\Lambda\over 2 m_b}\,r_b(v\cdot p) + {\cal{O}}(1/m_Q^2) \,,
\end{equation}
where the function
\begin{equation}\label{rQ}
   r_Q(v\cdot p) = 1 + {1\over\bar\Lambda\,\hat B}\,\bigg(
   - v\cdot p\,\hat A + 4\hat F_6 - \hat G_2 - 2\hat G_3^{\rm reg}
   + 2\hat G_5 - 6\hat G_6 \bigg)
\end{equation}
depends logarithmically on $m_Q$ through the definition of the
renormalized form factors in appendix~\ref{app:1}. The function
$\hat G_3^{\rm reg}$ has been defined in (\ref{G3reg}). The accuracy
with which $|V_{ub}|$ can be determined depends crucially upon how well
one will be able to estimate the $1/m_Q$ corrections in (\ref{rQ}).
Thus, a detailed investigation of the leading ($\hat A$ and $\hat B$)
and subleading ($\hat F_6$ and $\hat G_i$) universal functions is most
desirable. Such an analysis is beyond the scope of the present paper.
We note, however, that for the related case of $B\to\rho\,\ell\,\nu$
decays, the form factor ratio corresponding to (\ref{RBDdef}) has been
calculated in the quark model, using a $1/m_Q$ expansion \cite{Dib}.
The results are encouraging in that the deviations from the flavor
symmetry limit turn out to be small, of order 15\%. We expect
corrections of similar size for the case of $B\to\pi$ transitions. In
fact, assuming that $r_Q$ is of order unity, we expect that the scale
of power corrections is set by
\begin{equation}
   {\bar\Lambda\over 2 m_c} - {\bar\Lambda\over 2 m_b}
   \simeq 11\% \,,
\end{equation}
where we have used $\bar\Lambda=0.5$ GeV, $m_c=1.5$ GeV, and $m_b=4.8$
GeV for the sake of argument.

Are there any indications that this estimate might be too optimistic?
We think not. The reason is that current algebra puts powerful
constraints on the form factors in the soft pion limit. In particular,
it fixes the normalization of $f_+$ at zero recoil. Using
(\ref{wellknown}) one obtains
\begin{equation}\label{RBDsoft}
   \lim_{p\to 0} R_{BD}(v\cdot p)
   = {g_{B B^*\pi}\over g_{D D^*\pi}}\,
   {f_{B^*}\sqrt{m_B}\over f_{D^*}\sqrt{m_D}} \,.
\end{equation}
While this relation was derived before in the infinite heavy quark mass
limit \cite{Wise,Burd,LW,HQS2}, we have shown that it is actually valid
to next-to-leading order in $1/m_Q$. It is well-known that, for
pseudoscalar mesons, there are substantial corrections to the
asymptotic scaling law $f_B\sqrt{m_B}\approx f_D\sqrt{m_D}$, which
enhance the ratio $f_B/f_D$. Theoretical predictions typically fall in
the range $(f_B\sqrt{m_B})/(f_D\sqrt{m_D})\simeq 1.3-1.5$
\cite{latt1,latt2,latt3,latt4,MN,BBBD,Elet,DoPa}. However, in
(\ref{RBDsoft}) there appear the decay constants of vector mesons. Both
QCD sum rules and lattice gauge theory predict that spin symmetry
violating corrections decrease the ratio $f_{B^*}/f_{D^*}$ as compared
to $f_B/f_D$. The predictions are $f_{B^*}/f_{D^*} = \kappa (f_B/f_D)$
with $\kappa=0.79\pm 0.03$ from QCD sum rules \cite{MN}, and
$\kappa=0.86\pm 0.06$ from lattice gauge theory \cite{latt2}. The total
effect is that the scaling violations are much smaller for vector meson
decay constants. One expects
\begin{equation}
   {f_{B^*}\sqrt{m_B}\over f_{D^*}\sqrt{m_D}} \simeq 1.05 - 1.20 \,,
\end{equation}
i.e., a rather moderate correction to the flavor symmetry limit.
Although we are not able to give a similar estimate for the ratio
$g_{B B^*\pi}/g_{D D^*\pi}$ in (\ref{RBDsoft}), we see no reason why it
should deviate from unity by an anomalously large amount. Hence, we
believe that the deviations from the symmetry prediction $R_{BD}=1$ are
of the naively expected order of magnitude. We conclude that from a
comparison of the lepton spectra in $B\to\pi\,\ell\,\nu$ and
$D\to\pi\,\ell\,\nu$ decays, it should be possible to extract
$|V_{ub}|$ in a model independent way with a theoretical uncertainty of
$10-20\%$. This would already be a major improvement over the current,
largely model dependent determination of $|V_{ub}|$ from inclusive
decays. To achieve an even higher precision, it is necessary to study
in detail the $1/m_Q$ corrections in (\ref{RBD}). It is only at this
level that hadronic uncertainties enter the analysis. In this paper, we
have developed the theoretical framework for such an investigation.

At this point it is necessary to discuss the validity of the various
expansions considered in this paper. The heavy quark expansion is valid
as long as, in the rest frame of the initial heavy meson, the energy of
the light degrees of freedom before and after the weak decay is small
compared to (twice) the heavy quark mass.\footnote{For a discussion of
the factor 2, see Ref.~\cite{review}.}
Hence, one must require that
\begin{equation}
   {\bar\Lambda\over 2 m_Q} \ll 1 \,, \qquad
   {v\cdot p\over 2 m_Q} \ll 1 \,,
\end{equation}
where $\bar\Lambda$ is the effective mass of the light degrees of
freedom in the initial heavy meson \cite{AMM}. The first ratio is of
order 5\% for $Q=b$ and 15\% for $Q=c$, whereas the second ratio varies
roughly between 0 and 1/4 for $m_\pi\le v\cdot p\le\case{1}/{2}
(m_B^2+m_\pi^2)/m_B$. Hence, we expect the heavy quark expansion to
hold over essentially the entire kinematic range accessible in
semileptonic decays. This assertion is in fact supported by quark model
calculations \cite{Nathan,Gust}.

Another important question is over what range in $v\cdot p$ can one
trust the leading term in the chiral expansion, i.e., the soft pion
relations given in (\ref{wellknown}) and (\ref{RBDsoft}). Since the
pion is a pseudo-Goldstone boson associated with the spontaneous
breaking of chiral symmetry, we expect that the scale for the momentum
dependence of the universal form factors of HQET is set by
$\Lambda_\chi = 4\pi f_\pi$, which is the characteristic scale of
chiral symmetry breaking. Although one should not take this naive
dimensional argument too seriously, we may argue that the universal
functions are slowly varying in $x=v\cdot p/\Lambda_\chi$, and the
leading chiral behavior should be a good approximation until
$v\cdot p\sim 1$ GeV. Hence, we expect that (\ref{wellknown}) and
(\ref{RBDsoft}) should not only hold near $v\cdot p=0$, but actually
over a rather wide range in $v\cdot p$. Recent QCD sum rule
calculations of the $q^2$-dependence of $f_+^{B\to\pi}(q^2)$ in the
full theory support this expectation. The authors of Ref.~\cite{SR4}
find that the pole formula (\ref{wellknown}) gives an excellent fit to
their theoretical calculation over the wide range $0\le q^2\le 20$
GeV$^2$. For the residue at $q^2=0$, they obtain $f_+^{B\to\pi}(0) =
0.26\pm 0.03$, which is consistent with other sum rule calculations
\cite{SR1,SR2,SR3,SR5,SR6}, and with the quark model prediction of
Ref.~\cite{BSW}.

This result is interesting since, by means of (\ref{wellknown}), the
residue can be translated into a value for the $B B^*\pi$ coupling
constant, yielding
\begin{equation}
   g_{B B^*\pi} \simeq 0.17\,
   \bigg( {200~{\rm MeV}\over f_{B^*}} \bigg) \,. \qquad
   (\hbox{\rm from Ref.~\cite{SR4}})
\end{equation}
This value is significantly smaller than a naive estimate based on PCAC
and the nonrelativistic constituent quark model, which gives
$g_{B B^*\pi}\simeq 1$ \cite{Nuss,IWis,Yan}. However, it has been
pointed out that this number may indeed be too large. From a
generalization of the Nambu--Jona-Lasinio model, the authors of
Ref.~\cite{Hill} find $g_{B B^*\pi}\simeq 0.32$. The large spread in
the theoretical predictions for the $B B^*\pi$ coupling constant poses
the question whether it is possible to obtain experimental information
on this coupling. So far, attempts in this direction have focussed on
the decays of charm mesons, assuming heavy quark symmetry (i.e.,
neglecting $1/m_c$ corrections). From the width of the $D^{*+}$, one
can derive the rather loose upper bound $g_{D D^*\pi}<1.7$
\cite{Wise}.\footnote{The tighter bound $g_{D D^*\pi}<0.7$ is obtained
when one uses the value $\Gamma(D^{*+})<131$ keV reported in
Ref.~\cite{ACCMOR}.}
The analysis of radiative $D^*$ decays in Refs.~\cite{Cho,Jim} allows
$0<g_{D D^*\pi}<1$. Finally, one can combine the measured branching
ratio for $D^0\to\pi^- e^+\nu$ with the assumption of a monopole
behavior of the form factor $f_+^{D\to\pi} (q^2)$ to obtain $g_{D
D^*\pi}\simeq (0.40\pm 0.15)\times (200~{\rm MeV}/f_{D^*})$
\cite{Casa,Flei}. All these determinations have large uncertainties,
however. The semileptonic decay $B\to\pi\,\tau\,\nu$, on the other
hand, offers a rather clean measurement of $g_{B B^*\pi}$. By measuring
the distribution in the decay angle between the pion and the lepton, it
is possible to disentangle the contributions of the form factors $f_+$
and $f_0$ to the decay rate \cite{Korn}. By means of (\ref{wellknown}),
such a measurement would determine the ratio $g_{B B^*\pi}
(f_{B^*}/f_B)\simeq 1.1\,g_{B B^*\pi}$, where we have used the results
of Refs.~\cite{latt2,MN} for the ratio of decay constants. This might
be one of the best ways to determine this important coupling constant
experimentally.

\acknowledgments
G.B.\ thanks John Donoghue for helpful discussions and suggestions.
M.N. thanks Mark Wise for useful discussions. Z.L.\ and M.N.\ wish to
thank the organizers of TASI-93 for the stimulating atmosphere and the
relaxing ambience, which helped to complete parts of this work. G.B.\
acknowledges partial support from the U.S.~National Science Foundation.
M.N.\ gratefully acknowledges financial support from the BASF
Aktiengesellschaft and from the German National Scholarship Foundation.
Y.N.\ is an incumbent of the Ruth E. Recu Career Development chair, and
is supported in part by the Israel Commission for Basic Research, by
the United States -- Israel Binational Science Foundation (BSF), and by
the Minerva Foundation. This work was also supported by the Department
of Energy, contract DE-AC03-76SF00515.

\appendix
\section{Radiative corrections}
\label{app:1}

The renormalization scale dependence of the universal functions of HQET
can be derived from the requirement that the physical meson form factors
defined in (\ref{ffdef}) and (\ref{hidef}) be $\mu$-independent.
Using the explicit expressions for the Wilson coefficients given in
(\ref{Cmag}) and (\ref{Wilson}), we find that in leading logarithmic
approximation the following combinations of functions are
renormalization-group invariant:
\begin{eqnarray}\label{zdef}
   z_1(v\cdot p) &=& F_5(v\cdot p,\mu) + v\cdot p\,B(v\cdot p,\mu)
    \,, \nonumber\\
   z_2(v\cdot p) &=& F_6(v\cdot p,\mu) + {1\over3}
    (\bar\Lambda - v\cdot p)\,B(v\cdot p,\mu) \,, \nonumber\\
   z_3(v\cdot p) &=& {G_1(v\cdot p,\mu)\over A(v\cdot p,\mu)}
    - {16\over3\beta}\,(\bar\Lambda - v\cdot p)\,\ln[\alpha_s(\mu)]
    \,, \nonumber\\
   z_4(v\cdot p) &=& {G_2(v\cdot p,\mu)\over B(v\cdot p,\mu)}
    - {16\over3\beta}\,(\bar\Lambda - v\cdot p)\,\ln[\alpha_s(\mu)]
    \,, \nonumber\\
   z_5(v\cdot p) &=& [\alpha_s(\mu)]^{-1/\beta}\,
    \bigg\{ G_3(v\cdot p,\mu) - G_5(v\cdot p,\mu) + G_6(v\cdot p,\mu)
    + {8\over9}(\bar\Lambda - v\cdot p)\,B(v\cdot p,\mu) \bigg\}
    \,, \nonumber\\
   z_6(v\cdot p) &=& [\alpha_s(\mu)]^{-1/\beta}\,
    \bigg\{ G_4(v\cdot p,\mu) - {1\over3}(1-\widehat p^2)\,
    G_5(v\cdot p,\mu) + G_6(v\cdot p,\mu) \nonumber\\
   &&\phantom{ [\alpha_s(\mu)]^{-1/\beta}\bigg\{ }
    \mbox{}- {8\over27}(\bar\Lambda - v\cdot p)\,\Big[
    A(v\cdot p,\mu) + B(v\cdot p,\mu) \Big] \bigg\} \,, \nonumber\\
   z_7(v\cdot p) &=& [\alpha_s(\mu)]^{-1/\beta}\,
    \bigg\{ G_5(v\cdot p,\mu) - {2\over3}\,F_5(v\cdot p,\mu)
    - {2\over 3}\,v\cdot p\,B(v\cdot p,\mu) \bigg\} \,, \nonumber\\
   z_8(v\cdot p) &=& [\alpha_s(\mu)]^{-1/\beta}\,
    \bigg\{ G_6(v\cdot p,\mu) + {2\over3}\,F_6(v\cdot p,\mu)
    - {2\over27}(\bar\Lambda - v\cdot p)\,B(v\cdot p,\mu) \bigg\} \,.
\end{eqnarray}
These renormalized functions are still universal in that they do not
depend on the heavy quark mass. In the next step, we define related
renormalized functions $\hat F_i(v\cdot p)$ and $\hat G_i(v\cdot p)$ in
analogy to (\ref{Ahatdef}), by requiring that they be $\mu$-independent
and agree at tree-level with the original functions $F_i$ and $G_i$.
This necessarily introduces logarithmic dependence on the heavy quark
mass. We obtain, again, in leading logarithmic approximation:
\begin{eqnarray}\label{ffrendef}
   \hat F_5(v\cdot p) &=& F_5(v\cdot p,\mu) - \Big(
    x^{2/\beta} - 1 \Big)\,v\cdot p\,B(v\cdot p,\mu) \,, \nonumber\\
   \hat F_6(v\cdot p) &=& F_6(v\cdot p,\mu) - {1\over3}\,
    \Big( x^{2/\beta} - 1 \Big)\,(\bar\Lambda - v\cdot p)\,
    B(v\cdot p,\mu) \,, \nonumber\\
   {\hat G_1(v\cdot p)\over\hat A(v\cdot p)} &=&
    {G_1(v\cdot p,\mu)\over A(v\cdot p,\mu)} - {16\over3\beta}\,
    (\bar\Lambda - v\cdot p)\,\ln x \,, \nonumber\\
   {\hat G_2(v\cdot p)\over\hat B(v\cdot p)} &=&
    {G_2(v\cdot p,\mu)\over B(v\cdot p,\mu)} - {16\over3\beta}\,
    (\bar\Lambda - v\cdot p)\,\ln x \,, \nonumber\\
   \hat G_3(v\cdot p) &=& x^{-1/\beta}\,G_3(v\cdot p,\mu)
    - {32\over27}\,\Big( x^{2/\beta} - x^{-1/\beta} \Big)\,
    (\bar\Lambda - v\cdot p)\,B(v\cdot p,\mu) \nonumber\\
   &+& {2\over3}\,\Big( 1 - x^{-1/\beta} \Big)\,\Big[
    F_5(v\cdot p,\mu) + F_6(v\cdot p,\mu) + \case{1}/{3}
    (\bar\Lambda + 2 v\cdot p)\,B(v\cdot p,\mu) \Big] \,,
    \nonumber\\
   \hat G_4(v\cdot p) &=& x^{-1/\beta}\,G_4(v\cdot p,\mu)
    + {8\over27}\,\Big( x^{2/\beta} - x^{-1/\beta} \Big)\,
    (\bar\Lambda - v\cdot p)\,A(v\cdot p,\mu) \nonumber\\
   &+& {2\over9}\,\Big( 1 - x^{-1/\beta} \Big)\,\Big[
    (1-\widehat p^2)\,F_5(v\cdot p,\mu) + 3 F_6(v\cdot p,\mu)
    + (\bar\Lambda - v\cdot p\,\widehat p^2)\,B(v\cdot p,\mu) \Big]
    \,, \nonumber\\
   \hat G_5(v\cdot p) &=& x^{-1/\beta}\,G_5(v\cdot p,\mu)
    + {2\over3}\,\Big(1 - x^{-1/\beta} \Big)\,\big[
    F_5(v\cdot p,\mu) + v\cdot p\,B(v\cdot p,\mu)] \,, \nonumber\\
   \hat G_6(v\cdot p) &=& x^{-1/\beta}\,G_6(v\cdot p,\mu)
    + {8\over27}\,\Big(x^{2/\beta} - x^{-1/\beta} \Big)\,
    (\bar\Lambda - v\cdot p)\,B(v\cdot p,\mu) \nonumber\\
   &-& {2\over3}\,\Big( 1 - x^{-1/\beta} \Big)\,\Big[
    F_6(v\cdot p,\mu) + \case{1}/{3}(\bar\Lambda-v\cdot p)\,
    B(v\cdot p,\mu) \Big] \,,
\end{eqnarray}
where $x=\alpha_s(\mu)/\alpha_s(m_Q)$. Using (\ref{Ahatdef}) and
(\ref{zdef}), it is readily seen that these functions are indeed
$\mu$-independent. In terms of them, the $1/m_Q$ expansion of any meson
form factor assumes the same form as at tree-level.

\section{Soft pion relations for $\hat G_{\lowercase{i}}
(\lowercase{v}\cdot\lowercase{p})$}
\label{app:2}

In this appendix we derive the soft pion relations for the subleading
form factors $G_i$, which arise from insertions of the $1/m_Q$
corrections in the effective Lagrangian into matrix elements of the
leading-order currents. The corresponding corrections to meson decay
constants are \cite{MN}\footnote{The constants ${\cal{G}}_i(\mu)$ were
denoted by $G_i(\mu)$ in the original paper.}
\begin{eqnarray}\label{deconst}
   \langle\,0\,|\,i\int{\rm d}y\,T\big\{ \bar q\,\Gamma\,h_v(0),
    O_{\rm kin}(y) \big\}\,|M(v)\rangle &=& i F(\mu)\,
    {\cal{G}}_1(\mu)\,{\rm Tr}\big\{ \Gamma\,{\cal{M}}(v) \big\} \,,
    \nonumber\\
   && \\
   \langle\,0\,|\,i\int{\rm d}y\,T\big\{ \bar q\,\Gamma\,h_v(0),
    O_{\rm mag}(y) \big\}\,|M(v)\rangle &=& 2i d_M F(\mu)\,
    {\cal{G}}_2(\mu)\,{\rm Tr}\big\{ \Gamma\,{\cal{M}}(v) \big\} \,,
    \nonumber
\end{eqnarray}
where $d_M=3$ for a pseudoscalar meson, and $d_M=-1$ for a vector
meson. The corrections to the coupling of two heavy mesons to the axial
vector current can be written as
\begin{eqnarray}
   \langle M'(v,p)|\,i\int{\rm d}y\,T\big\{ \,p\cdot A(0),
    O_{\rm kin}(y) \big\}\,|M(v)\rangle &=& g_1\,{\rm Tr}\Big\{
    \gamma_5\,\,\rlap/\!p\,\overline{\cal{M}}'(v){\cal{M}}(v) \Big\}
    + \ldots \,, \nonumber\\
   && \\
   \langle M'(v,p)|\,i\int{\rm d}y\,T\big\{ \,p\cdot A(0),
    O_{\rm mag}(y) \big\}\,|M(v)\rangle &=& 2(d_M + d_{M'})\,g_2(\mu)
    \,{\rm Tr}\Big\{ \gamma_5\,\,\rlap/\!p\,\overline{\cal{M}}'(v)
    {\cal{M}}(v) \Big\} \nonumber\\
   &&\mbox{}+ \ldots \,, \nonumber
\end{eqnarray}
where the ellipses denote terms quadratic and higher order in $p$.

Let us first work out the pole terms arising from an insertion of
$O_{\rm kin}$. According to Fig.~\ref{fig:2}a, there are two single
pole contributions:
\begin{eqnarray}
   \sum_{M'} &&\bigg[ \langle\,0\,|\,\bar q\,\Gamma\,h_v\,|M'(v)\rangle
    \,{i\over 2 v\cdot(-p)}\,\langle M'(v)|\,i\int{\rm d}y\,T\big\{
    \,p\cdot A(0),O_{\rm kin}(y) \big\}\,|M(v)\rangle \nonumber\\
   &&\mbox{}+ \langle\,0\,|\,i\int{\rm d}y\,T\big\{ \bar q\,\Gamma\,
    h_v(0),O_{\rm kin}(y) \big\}\,|M'(v)\rangle\,{i\over 2 v\cdot(-p)}
    \,\langle M'(v)|\,p\cdot A\,|M(v)\rangle \bigg] \nonumber\\
   =&&\mbox{}- {F(\mu)\over 2 v\cdot p}\,\Big[ 2 g\,{\cal{G}}_1(\mu)
    + g_1 \Big]\,{\rm Tr}\Big\{ \gamma_5\,(\,\rlap/\!p - v\cdot p)\,
    \Gamma\,{\cal{M}}(v) \Big\} + \ldots \,.
\end{eqnarray}
As shown in Fig.~\ref{fig:2}b, there are also potential double pole
contributions. The first diagram gives rise to
\begin{eqnarray}\label{dpole}
   &&\sum_{M',M''}\,\langle\,0\,|\,\bar q\,\Gamma\,h_v\,|M''(v)\rangle
    \,\langle M''(v)|\,O_{\rm kin}\,|M'(v)\rangle\,
    \langle M'(v)|\,p\cdot A\,|M(v)\rangle\,
    \bigg( {i\over 2 v\cdot(-p)} \bigg)^2 \nonumber\\
   &&\quad = - {\lambda_1\over 2(v\cdot p)^2}\,
    \sum_{M'}\,\langle\,0\,|\,\bar q\,\Gamma\,h_v\,|M'(v)\rangle\,
    \langle M'(v)|\,p\cdot A\,|M(v)\rangle \,.
\end{eqnarray}
Note that only diagonal terms ($M''=M'$) contribute to the sum. We have
introduced the mass parameter $\lambda_1$, which parameterizes the
matrix element of the kinetic operator. In general, one defines
\begin{eqnarray}\label{lambdef}
   \langle M(v)|\,O_{\rm kin}\,|M(v)\rangle &=& 2\,\lambda_1 \,,
    \nonumber\\
   \langle M(v)|\,O_{\rm mag}\,|M(v)\rangle &=& 2 d_M\,\lambda_2(\mu)
    \,.
\end{eqnarray}
The same matrix elements also determine the $1/m_Q$ corrections to the
physical meson masses:
\begin{equation}\label{masses}
   m_M = m_Q + \bar\Lambda - {1\over 2 m_Q}\,\Big[ \lambda_1
   + d_M\,C_{\rm mag}(\mu)\,\lambda_2(\mu) \Big] + \ldots \,.
\end{equation}
This induces a mass renormalization which modifies the meson
propagator, as shown in the second diagram in Fig.~\ref{fig:2}b. The
corresponding correction is obtained from the expansion
\begin{equation}
   {i M\over[(M + \varepsilon) v + k]^2 - M^2}
   \stackrel{M\to\infty}{=} {i\over 2 v\cdot k}\,
   \bigg( 1 - {\varepsilon\over v\cdot k} + \ldots \bigg) \,,
\end{equation}
where $M=m_Q+\bar\Lambda$. For the kinetic operator,
$\varepsilon=-\lambda_1/2 m_Q$. (The $\lambda_2$ term will be taken
into account below.) Combining this with the leading-order pole
contribution in (\ref{LOpole}), we find that the contribution from mass
renormalization exactly cancels the double pole contribution
(\ref{dpole}). As a result, only the commutator and the single pole
terms remain, and we obtain the soft pion relations
\begin{eqnarray}\label{softG12}
   G_1(0,\mu) &=& {F(\mu)\over 2 f_\pi}\,\Big[ 2(1-g)\,
    {\cal{G}}_1(\mu) - g_1 \Big] \,, \nonumber\\
   G_2(0,\mu) &=& {F(\mu)\over 2 f_\pi}\,\Big[ 2 g\,{\cal{G}}_1(\mu)
    + g_1 \Big] \,.
\end{eqnarray}

Things are slightly more complicated in the case of an insertion of the
chromo-magnetic operator $O_{\rm mag}$. The single pole contributions
are
\begin{equation}\label{last}
   {F(\mu)\over 2 v\cdot p}\,\sum_{M'}\,\Big[ 2 d_{M'}\,g\,
   {\cal{G}}_2(\mu) + (d_M+d_{M'})\,g_2(\mu) \Big]\,{\rm Tr}\big\{
   \Gamma\,{\cal{M}}'(v) \big\}\,{\rm Tr}\Big\{ \gamma_5\,\,\rlap/\!p\,
   \overline{\cal{M}}'(v){\cal{M}}(v) \Big\} \,.
\end{equation}
To recover the trace structures appearing in (\ref{Gdef}), we need a
second trace identity besides (\ref{magictrace}). It is
\begin{equation}
   \sum_{M'}\,(d_{M'}-d_M)\,{\rm Tr}\big\{ \Gamma\,{\cal{M}}'(v)
   \big\}\,{\rm Tr}\Big\{ \gamma_5\,\,\rlap/\!p\,
   \overline{\cal{M}}'(v){\cal{M}}(v) \Big\}
   = -4\,{\rm Tr}\bigg\{ i\gamma_5\,p_\alpha\gamma_\beta\,\Gamma\,
   {1+\rlap/v\over 2}\,\sigma^{\alpha\beta} {\cal{M}}(v) \bigg\} \,.
\end{equation}
This allows us to rewrite our result (\ref{last}) as
\begin{eqnarray}
  &&-{F(\mu)\over v\cdot p}\,\Big[ g\,{\cal{G}}_2(\mu) + g_2(\mu)
   \Big]\,{\rm Tr}\bigg\{ \gamma_5\,\sigma_{\alpha\beta}\,
   (\,\rlap/\!p - v\cdot p)\,\Gamma\,{1+\rlap/v\over 2}\,
   \sigma^{\alpha\beta} {\cal{M}}(v) \bigg\} \nonumber\\
  &&-{2 F(\mu)\over v\cdot p}\,\Big[ 2 g\,{\cal{G}}_2(\mu) + g_2(\mu)
   \Big]\,{\rm Tr}\bigg\{ i\gamma_5\,p_\alpha\gamma_\beta\,\Gamma\,
   {1+\rlap/v\over 2}\,\sigma^{\alpha\beta} {\cal{M}}(v) \bigg\} \,,
\end{eqnarray}
where we have used that $\case{1}{2}(1+\rlap/v)\,\sigma_{\alpha\beta}
{\cal{M}}(v)\,\sigma^{\alpha\beta} = 2 d_M\, {\cal{M}}(v)$ \cite{MN}.
The double pole contribution can be calculated in complete analogy to
the case of the kinetic operator, except that the contribution from
mass renormalization will not cancel the direct double pole term, since
the spin of the pole meson can be different from the spin of the
external heavy meson. In fact, we find
\begin{eqnarray}
   &&-{\lambda_2\over 2(v\cdot p)^2}\,\sum_{M'}\,(d_{M'}-d_M)\,
    \langle\,0\,|\,\bar q\,\Gamma\,h_v\,|M'(v)\rangle\,
    \langle M'(v)|\,p\cdot A\,|M(v)\rangle \nonumber\\
  &&\qquad = i F(\mu)\,\lambda_2(\mu)\,{g(v\cdot p)\over(v\cdot p)^2}\,
   {\rm Tr}\bigg\{ \gamma_5\,i p_\alpha\gamma_\beta\,\Gamma\,
   {1+\rlap/v\over 2}\,\sigma^{\alpha\beta}\,{\cal{M}}(v) \bigg\} \,.
\end{eqnarray}
The parameter $\lambda_2(\mu)$ has been defined in (\ref{lambdef}).
Collecting the commutator, single and double pole contributions and
comparing the result with (\ref{Gdef}), we find the following soft pion
relations:
\begin{eqnarray}\label{softG36}
   \lim_{p\to 0} G_3(v\cdot p) &=& {F(\mu)\over 2 f_\pi}\,\bigg[
    - 2\lambda_2(\mu)\,{g(v\cdot p)\over v\cdot p}
    - 8 g\,{\cal{G}}_2(\mu) - 4 g_2(\mu) \bigg] \,, \nonumber\\
   G_4(0) &=& {F(\mu)\over 2 f_\pi}\,\Big[ 2 (1-g)\,{\cal{G}}_2(\mu)
    - 2 g_2(\mu) \Big] \,, \nonumber\\
   G_5(0) &=& 0 \,, \nonumber\\
   G_6(0) &=& {F(\mu)\over 2 f_\pi}\,\Big[ 2 g\,{\cal{G}}_2v
    + 2 g_2(\mu) \Big] \,.
\end{eqnarray}

To obtain the corresponding relations for the renormalized functions
$\hat G_i(v\cdot p)$, one first has to renormalize the low-energy
parameters appearing on the right-hand side of the soft pion relations.
It is easy to see that $g$, $g_1$, and $\lambda_1$ are not renormalized
to all orders in perturbation theory, whereas the $\mu$-dependence of
$g_2(\mu)$ and $\lambda_2(\mu)$ is compensated by the Wilson
coefficient of the chromo-magnetic operator. The renormalization of
${\cal G}_{1,2}$ is slightly more complicated. It is discussed in
Ref.~\cite{MN}. In leading logarithmic approximation, the renormalized
low-energy parameters are given by
\begin{eqnarray}\label{stuff}
   \phantom{ \bigg[ }
   \hat g_2 &=& x^{-3/\beta}\,\hat g_2(\mu) \,, \qquad
    \hat \lambda_2 = x^{-3/\beta}\,\lambda_2(\mu) \,, \nonumber\\
   \hat{\cal G}_1 &=& {\cal G}_1(\mu) - {8\bar\Lambda\over 3\beta}\,
    \ln x \,, \qquad
    \hat{\cal G}_2 = x^{-3/\beta}\,\bigg[ {\cal G}_2(\mu)
    - {4\bar\Lambda\over 27} \bigg] \,,
\end{eqnarray}
where $x=\alpha_s(\mu)/\alpha_s(m_Q)$. By means of (\ref{masses}),
$\hat\lambda_2$ is related to the mass splitting between vector and
pseudoscalar mesons:
\begin{equation}
   \hat\lambda_2 = {1\over 4}\,(m_V^2 - m_P^2) \,.
\end{equation}
The soft pion relations (\ref{softG}) and (\ref{softG3}) follow
by combining (\ref{softG12}), (\ref{softG36}), (\ref{stuff}), and
(\ref{ffrendef}).

\begin{figure}
\caption{
Pole diagram contributing to the soft pion relations for the universal
form factors. The axial vector current is shown as the dashed line,
whereas the weak current is drawn as a wiggly line. The black dot
represents the strong interaction vertex, the open box the weak
interaction matrix element.}
\label{fig:1}
\end{figure}

\begin{figure}
\caption{
Single (a) and double (b) pole diagrams contributing to the soft pion
relations for the form factors $G_i$. The notation is the same as in
Fig.~1. In addition, a cross represents a $1/m_Q$ insertion of
$O_{\rm kin}$ or $O_{\rm mag}$.}
\label{fig:2}
\end{figure}

\end{document}